\documentclass[aps,prl,twocolumn,letterpaper,superscriptaddress,showpacs,nobibnotes]{revtex4-2}
\usepackage{keyval}%
\usepackage{graphicx}
\usepackage{dcolumn}
\usepackage{bm}
\usepackage{color}
\usepackage{amsmath,amssymb}

\begin{document}

\title{Stabilizing charge density wave by mixing transition metal elements in monolayer XS$_2$ with trigonal-prismatic coordination}
\author{Chi-Cheng Lee}
\affiliation{Department of Physics, Tamkang University, Tamsui, New Taipei 251301, Taiwan}%
\author{Yukiko Yamada-Takamura}
\affiliation{School of Materials Science, Japan Advanced Institute of Science and Technology (JAIST), 1-1 Asahidai, Nomi, Ishikawa 923-1292, Japan}%
\date{\today}

\begin{abstract}
The electronic structure and phonon dispersion of XS$_2$ with X = Co, Tc, Ti, Ru, Nb, and Rh in the monolayer MoS$_2$ structure with trigonal-prismatic coordination
are studied from first principles. 
Although each XS$_2$ is dynamically unstable, CoS$_2$, TcS$_2$, RuS$_2$, and RhS$_2$ can be stabilized by developing charge density waves in the (2$\times$2) supercell, 
leading to metal-insulator transitions. Without really needing the metal-insulator transitions and large atomic distortions,
additional energy may be gained in the total energy by mixing transition metal elements to create high-entropy combinations for X,
presenting a wide range of high-entropy XS$_2$ compounds that exhibit a variety of band structures, including direct- and indirect-gap semiconductors, metals, and semimetals.
\end{abstract}
  
\maketitle

The high-entropy alloy, also known as the multi-component alloy, is a single-phase solid-solution mixture of metal elements, with each element contributing a significant 
percentage to the composition\cite{HEA1,HEA2,HEA3,HEA4,HEA5}. Most high-entropy alloys tend to crystallize in simple FCC, BCC, and HCP structures, 
similar to typical metal structures. The high-entropy alloy gains energy from the high configurational entropy of mixing by lowering 
the Gibbs free energy\cite{Yeh,PhysRevLett.113.107001}. The pioneering research conducted by Yeh \textit{et al.}\cite{Yeh} and Cantor \textit{et al.}\cite{Cantor} 
has shown that highly mixed equiatomic alloys can exhibit exceptional mechanical strength with excellent ductility, 
surpassing the fracture toughness achievable in conventional alloys\cite{HEA1,HEA2,HEA3,HEA4,HEA5}. 
Specifically, Cantor \textit{et al.} fabricated a high-entropy alloy consisting of a five-component alloy, FeCrMnNiCo, and Yeh \textit{et al.} synthesized an equimolar 
mixture of CuCoNiCrAl$_x$Fe with varying Al concentrations. The high-entropy alloys have also shown promising performance in electrocatalysis\cite{Junlin} and can exhibit 
superconductivity, with the potential to further increase superconducting transition 
temperatures through proper mixing of various types of metal elements\cite{PhysRevLett.113.107001,PhysRevMaterials.3.060602,PhysRevMaterials.3.090301}.

\begin{figure}[tbp]
\includegraphics[width=1.00\columnwidth,clip=true,angle=0]{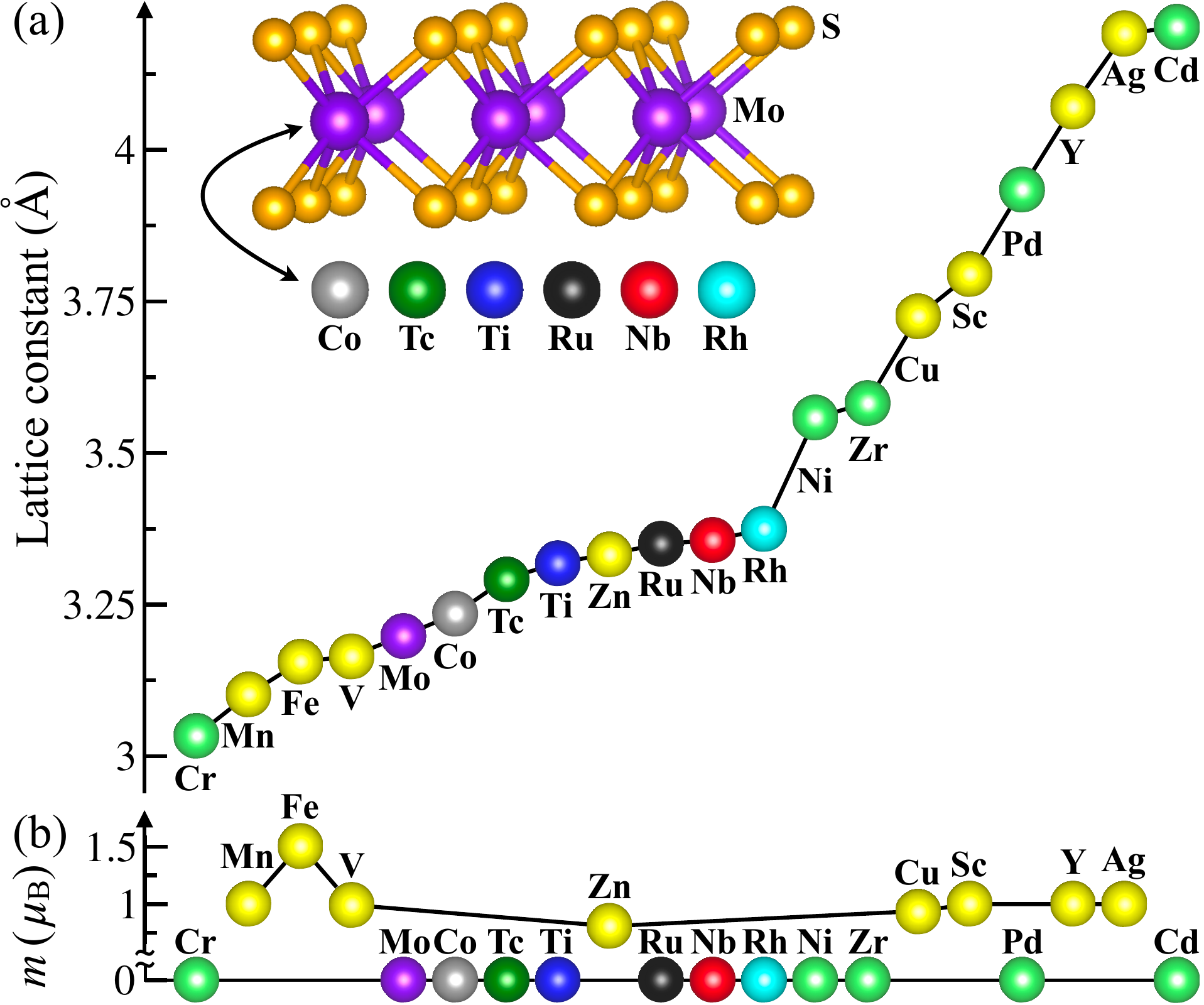}
\caption{(a) Lattice constants of monolayer XS$_2$ with X = Mo, Co, Tc, Ti, Ru, Nb, and Rh in the 1H-MoS$_2$ structure. 
(b) Total spin moment ($m$) of XS$_2$. 
}
\label{fig:Fig1}
\end{figure}

Contrary to the expectation that the entropy of mixing is the primary factor in stabilizing high-entropy alloys, 
Senkov \textit{et al.}\cite{HEA3} found that simply increasing the configurational 
entropy doesn't guarantee stability in high-entropy alloys after analyzing over 130,000 alloys using the calculated phase diagram method, 
suggesting that it's crucial to understand the role played by the total energy responsible for determining 
the formation enthalpy. First-principles approaches based on density functional theory\cite{Kohn,Sham} have also been applied to study magnetic properties, phase stability,
elastic properties, and local lattice distortions of high-entropy alloys\cite{Fuyang,IKEDA2019464,PhysRevLett.126.025501}. 
Samolyuk \textit{et al.}\cite{PhysRevLett.126.025501} argued that 
configurational entropy has minimal impact on stabilizing one crystal structure over another and the total energy gain from atomic displacement may be more 
important in eliminating dynamic instabilities in high-entropy alloys. Recently, two-dimensional van der Waals materials with a variety of possible heterostructures 
through stacking and twisting have become popular and promising for next-generation technology\cite{ref1,ref2,ref3,ref4}. 
High-entropy van der Waals compounds have also been extensively 
researched and produced experimentally, including metal dichalcogenides, halides and phosphorus trisulphides\cite{Zhiguo,nemani,Hosono1,Hosono2,RanWang,Zhaosu}.
Among the layered compounds, MoS$_2$ is an interesting semiconductor that undergoes an indirect-direct gap transition as the structure changes from bulk to 
monolayer\cite{Andrea,KUMAR}, useful for electronic and optoelectronic applications. 
The monolayer MoS$_2$ has a trigonal-prismatic structure\cite{Miklos}, which will be simply dubbed as 1H-MoS$_2$.
It is then pertinent to ask whether highly mixed transition metal elements formed in the ``1H-MoS$_2$'' structure 
can still be dynamically stable, able to engineer the band structure, and gain energy at the level of total energy without the assistance of 
the configurational entropy of mixing, given that most of the pristine 1H-XS$_2$ compounds, where X denotes the transition metal elements, are unstable 
unless charge density waves are developed in larger supercells\cite{Bianco,Knispel}.

In this study, we performed first-principles calculations within the generalized gradient approximation\cite{GGA} using the OpenMX code\cite{openmx}, 
where norm-conserving pseudopotentials and optimized pseudoatomic basis functions were adopted\cite{Ozaki}. Phonon dispersion was obtained based on 
supercell force-constant calculations, and the force constants were properly partitioned\cite{LEE}.
More details can be found in Supplemental Material\cite{Supplement}.
To investigate the potential two-dimensional high-entropy XS$_2$ compounds having the 1H-MoS$_2$ structure, we first relaxed the lattice 
parameters by considering X to be each of the transition metal elements in the fourth and fifth rows of the periodic table. 
The optimized lattice constants are depicted in Fig.~\ref{fig:Fig1} (a). The result indicates that the values of two lattice constants, 
for instance, 4.2022 \AA~for Cd and 3.0345 \AA~for Cr, could differ by more than one angstrom.
Even though a phase containing a mixed combination of all the explored transition metal elements could be stable, 
a good choice is to select the elements giving similar lattice constants, which can help maintain a structure that still closely 
resembles the 1H-MoS$_2$ structure. While MoS$_2$ does not exhibit magnetism, we have discovered that several XS$_2$ compounds 
can be magnetic, such as MnS$_2$, FeS$_2$, and VS$_2$. 
The total spin moments calculated for these pristine XS$_2$ compounds are shown in Fig.~\ref{fig:Fig1} (b).
To better understand energy gain without considering the influence of magnetism, we will focus on non-magnetic XS$_2$ compounds with similar lattice constants.
Specifically, we will examine the properties of XS$_2$ compounds with X = Mo, Co, Tc, Ti, Ru, Nb, and Rh in the following paragraphs.

\begin{figure}[tbp]
\includegraphics[width=1.00\columnwidth,clip=true,angle=0]{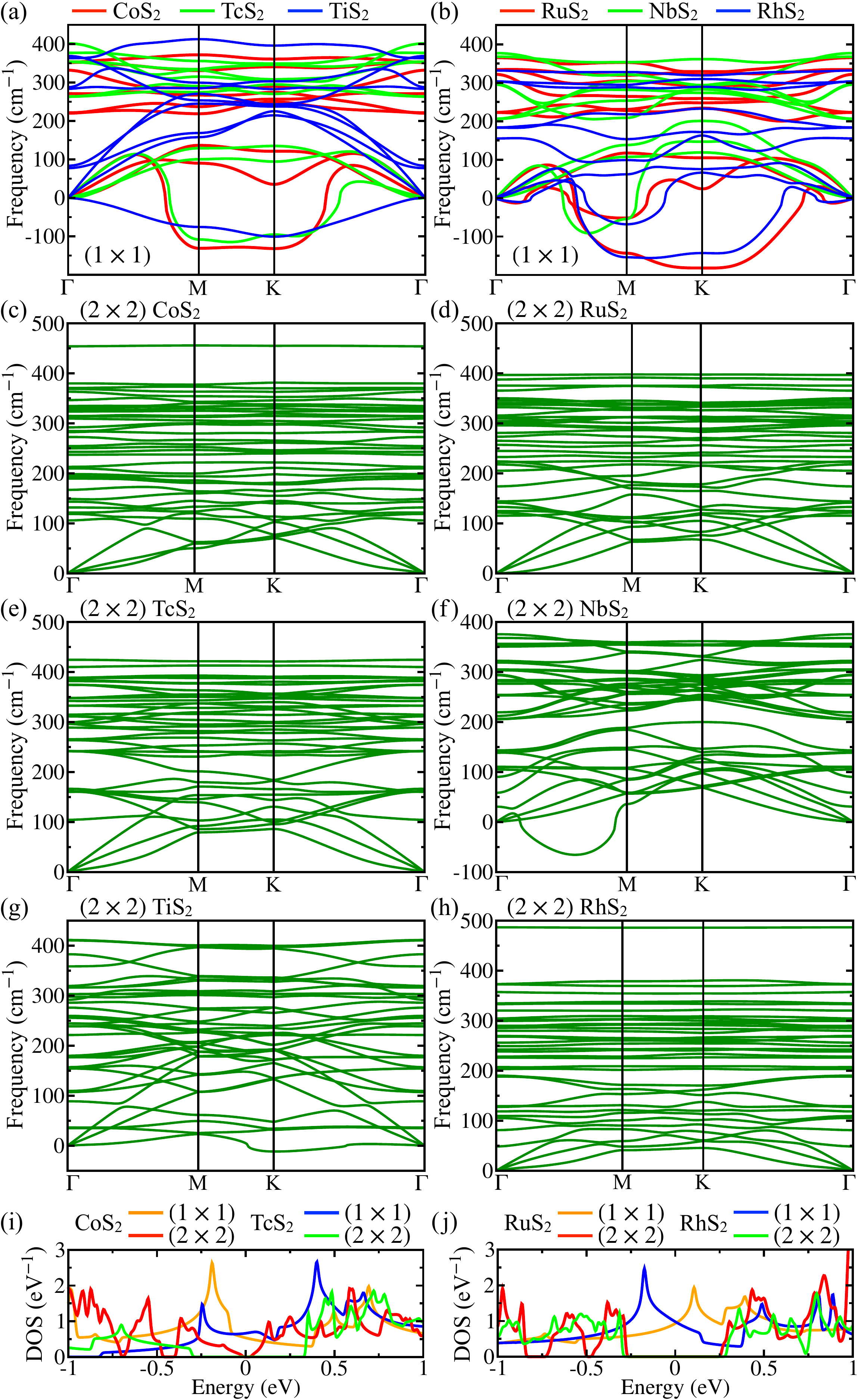}
\caption{Phonon dispersions of primitive-cell (1$\times$1) (a) CoS$_2$, TcS$_2$, and TiS$_2$ and (b) RuS$_2$, NbS$_2$, and RhS$_2$.
All the cases possess imaginary frequencies, which are presented by negative values.
The phonon dispersions of the charge-density-wave structures of (c) CoS$_2$, (d) RuS$_2$, (e) TcS$_2$, (f) NbS$_2$, (g) TiS$_2$, and (h) RhS$_2$
obtained in the (2$\times$2) supercell. Electronic density of states (DOS) of (i) CoS$_2$ and TcS$_2$ 
and (j) RuS$_2$ and RhS$_2$ in units of eV$^{-1}$ per atom. The Fermi level is shifted to 0 eV.
}
\label{fig:Fig2}
\end{figure}

The XS$_2$ compounds with X = Co, Tc, Ti, Ru, Nb, and Rh have been less studied in the 1H-MoS$_2$ structure. This is because of the 
existing dynamical instability, which can be seen in the phonon dispersions with imaginary frequencies, as demonstrated in 
Figs.~\ref{fig:Fig2} (a) and (b). The phonon instability is evident at the M point in all six compounds, indicating that charge density waves 
could be developed to stabilize the systems in the (2$\times$2) supercell. As shown in Figs.~\ref{fig:Fig2} (c) to (h), 
CoS$_2$, RuS$_2$, TcS$_2$, and Rh$_2$ become dynamically stable in the (2$\times$2) supercell due to the developed charge density waves. 
However, this is not the case for TiS$_2$ and NbS$_2$. As listed in Table~\ref{table:totalenergy}, all six compounds can gain energy
in the (2$\times$2) supercell compared to the corresponding energies calculated from the (1$\times$1) unit cells. We have not observed the development of 
a charge density wave in MoS$_2$, which is consistent with the stable dynamical behavior in the phonon dispersion of (1$\times$1) 1H-MoS$_2$. 
Even with the development of charge density waves, the crystal structures of (2$\times$2) TcS$_2$, (2$\times$2) TiS$_2$, and (2$\times$2) NbS$_2$ can still 
remain hexagonal, in contrast to the case of (2$\times$2) CoS$_2$, (2$\times$2) RuS$_2$, and (2$\times$2) RhS$_2$, where both lattice and atomic distortions 
can be found. We want to emphasize that the small energy gain of 0.2 meV per atom in (2$\times$2) NbS$_2$ is genuine, as the imaginary frequencies along other paths 
can be stabilized. The more complete phonon dispersions of all the studied compounds can be found in Supplemental Material\cite{Supplement}.

\begin{table}[tbp]
\caption{The band-gap type of each monolayer XS$_2$ is listed together with the total energy difference from 
the corresponding energies of (1$\times$1) XS$_2$ and (2$\times$2) XS$_2$ with X = Mo, Co, Tc, Ti, Ru, Nb, and Rh
in units of eV per atom, denoted as E-(1$\times$1) and E-(2$\times$2), respectively. 
The lattice parameters can be found in Supplemental Material\cite{Supplement}.
}
\label{table:totalenergy}%
\begin{tabular}{cccc}
\hline\hline
Monolayer 1H-XS$_2$     & Gap type & E-(1$\times$1) & E-(2$\times$2) \\ \hline
(2$\times$2) MoS$_2$ & direct   &   -0.0000    &   0      \\
(2$\times$2) CoS$_2$ & direct   &   -0.0526    &   0      \\
(2$\times$2) TcS$_2$ & indirect &   -0.0625    &   0      \\
(2$\times$2) TiS$_2$ & indirect &   -0.0017    &   0      \\
(2$\times$2) RuS$_2$ & indirect &   -0.0841    &   0      \\
(2$\times$2) NbS$_2$ & metallic &   -0.0002    &   0      \\
(2$\times$2) RhS$_2$ & indirect &   -0.1584    &   0      \\ \hline
MoCoTcTiS$_8$ & indirect              &    -0.0341    &   -0.0049    \\
MoCoTiRuS$_8$ & metallic              &    -0.0507    &   -0.0161    \\
MoCoTiNbS$_8$ & indirect              &    -0.0015    &    0.0121    \\
MoTiRuNbS$_8$ & metal                 &    -0.1044    &   -0.0829    \\
MoTiNbRhS$_8$ & indirect              &    -0.0836    &   -0.0435    \\
MoRuNbRhS$_8$ & indirect              &    -0.0571    &    0.0036    \\
CoTcTiNbS$_8$ & metallic              &    -0.0831    &   -0.0539    \\
CoTcRuRhS$_8$ & metallic              &    -0.0638    &    0.0256    \\
TiRuNbRhS$_8$ & semimetal             &    -0.1703    &   -0.1092    \\ \hline
Mo$_5$CoTiRuRhS$_{18}$ &     direct   &    -0.0183    &    0.0147    \\
Mo$_3$CoTcTiRuNbRhS$_{18}$ & indirect &    -0.0791    &   -0.0391    \\ \hline\hline
\end{tabular}%
\end{table}

The charge density waves in (2$\times$2) XS$_2$ involve both atomic distortion and charge order. Based on Mulliken population 
analysis, the largest charge difference in the X atom for X = Co, Tc, Ti, Ru, Nb, and Rh is 0.028, 0.030, 0.000, 0.015, 0.009, and 0.067 electrons, 
respectively. The largest charge difference in the S atom with the same sequence is 0.077, 0.035, 0.008, 0.110, 0.001, and 0.051 electrons, respectively. 
The negligible charge transfer in TiS$_2$ and NbS$_2$ are consistent with the slight energy gain and phonon instability in the 
(2$\times$2) supercell. The important role played by the electronic structure in stabilizing CoS$_2$, RuS$_2$, TcS$_2$, and Rh$_2$ in the (2$\times$2) 
supercell can also be revealed in the density of states with the presence of a gap at the Fermi level. As demonstrated in Figs.~\ref{fig:Fig2} (i) and (j), 
CoS$_2$, TcS$_2$, RuS$_2$, and RhS$_2$ all encounter metal-insulator transitions. In contrast, TiS$_2$ is already an insulator in the (1$\times$1) unit cell, 
and (2$\times$2) NbS$_2$ remains metallic. This aligns with the finding that there is insufficient energy gain through the charge transfer and 
atomic distortion in these two compounds. This is because the band gap is already open in one case, while it is still not open in the other case.

The commonly considered energy gain of high-entropy alloys from mixing different elements in thermodynamics is through the TS term in the Gibbs free energy, 
G $=$ U $+$ PV $-$ TS, where U, P, V, T, and S denote the total energy, pressure, volume, temperature, and entropy, respectively.
Instead of exploring how to increase the disorder in contributing to the entropy, we intend to investigate the energy gain of highly mixed transition metal 
elements in the 1H-MoS$_2$ structure by examining the impact of mixing on the total energy, rather than focusing on increasing randomness 
to contribute to the entropy of mixing. The density functional theory is effective for delivering such total energy and it is generally not expected to gain 
total energy in a multi-element system with high disorder. Otherwise, predicting the ground-state structures would become more challenging if introducing 
higher-entropy combinations could easily lower the total energy and still keep the system dynamically stable. In the paragraph below, we will demonstrate that 
while the XS$_2$ compounds with X = Co, Tc, Ti, Ru, Nb, and Rh are unstable in the (1$\times$1) unit cell, they become dynamically stable when they are mixed in 
the (2$\times$2) supercell. Furthermore, in some combinations, the total energy gain can be greater than the sum of the individual energy gains of the corresponding
(2$\times$2) XS$_2$ compounds.

We have investigated 9 combinations involving 4 out of the 7 transition metal elements: Mo, Co, Tc, Ti, Ru, Nb, and Rh in the (2$\times$2) supercell. 
These combinations are MoCoTcTiS$_8$, MoCoTiRuS$_8$, MoCoTiNbS$_8$, MoTiRuNbS$_8$, MoTiNbRhS$_8$, MoRuNbRhS$_8$, CoTcTiNbS$_8$, CoTcRuRhS$_8$, and TiRuNbRhS$_8$. 
Despite the mixing, both MoTiNbRhS$_8$ and MoTiRuNbS$_8$ can still keep the lattice hexagonal, and the atomic distortions compared to the perfect (2$\times$2) 1H-MoS$_2$ 
structure are tiny. The structures of MoTiNbRhS$_8$, CoTcRuRhS$_8$, and TiRuNbRhS$_8$ are presented in Fig.~\ref{fig:Fig3} (a). 
The other structures can be found in Supplemental Material\cite{Supplement}. Note that even with the significant distortion of S atoms 
in CoTcRuRhS$_8$, the structure can still be recognized as closely related to the 1H-MoS$_2$ structure.
As shown in Table~\ref{table:totalenergy}, all 9 compounds possess total energies lower than the total-energy sum of the corresponding (1$\times$1) XS$_2$ compounds. 
By further examining the total energy difference from 
the total-energy sum of the corresponding (2$\times$2) XS$_2$ compounds, MoCoTcTiS$_8$, MoCoTiRuS$_8$, MoTiRuNbS$_8$, MoTiNbRhS$_8$, CoTcTiNbS$_8$, and TiRuNbRhS$_8$ 
can gain more energy than the individual XS$_2$ compounds in the (2$\times$2) supercell. This suggests that mixing different transition metal elements can 
more efficiently develop charge density waves than the individual XS$_2$ compounds in gaining energy at the level of total energy, without the assistance of 
the TS term in the Gibbs free energy.

It is interesting to further examine the dynamical stability of these 9 compounds, including MoCoTiNbS$_8$, MoRuNbRhS$_8$, and CoTcRuRhS$_8$, where the total energies 
are lower than the summed total energy of the related (1$\times$1) XS$_2$ compounds but the energy gain is less efficient compared to arranging the same elements 
in the (2$\times$2) supercell. We have found that all the 9 compounds are dynamically stable without the presence of imaginary frequencies in the phonon dispersion.
For illustration, the dispersions of MoTiNbRhS$_8$, CoTcRuRhS$_8$, and TiRuNbRhS$_8$ are shown in Figs.~\ref{fig:Fig3} (c)-(e), respectively. 
The other phonon disperions can be found in Supplemental Material\cite{Supplement}. Note that neither TiS$_2$ nor NbS$_2$ is dynamically stable in the (2$\times$2) supercell, 
and MoS$_2$ does not develop a charge density wave. However, the combination of MoTiNb with Co, Ru, or Rh can stabilize the phonon instability. The phonon instability can 
also be removed by combining TiNb with CoTc or RuRh. As presented in Table~\ref{table:totalenergy}, the metal-insulator transition is not really needed to stabilize 
these high-entropy compounds in general. This is especially true given that MoCoTiRuS$_8$, MoTiRuNbS$_8$, CoTcTiNbS$_8$, and CoTcRuRhS$_8$ are metallic.

\begin{figure}[tbp]
\includegraphics[width=1.00\columnwidth,clip=true,angle=0]{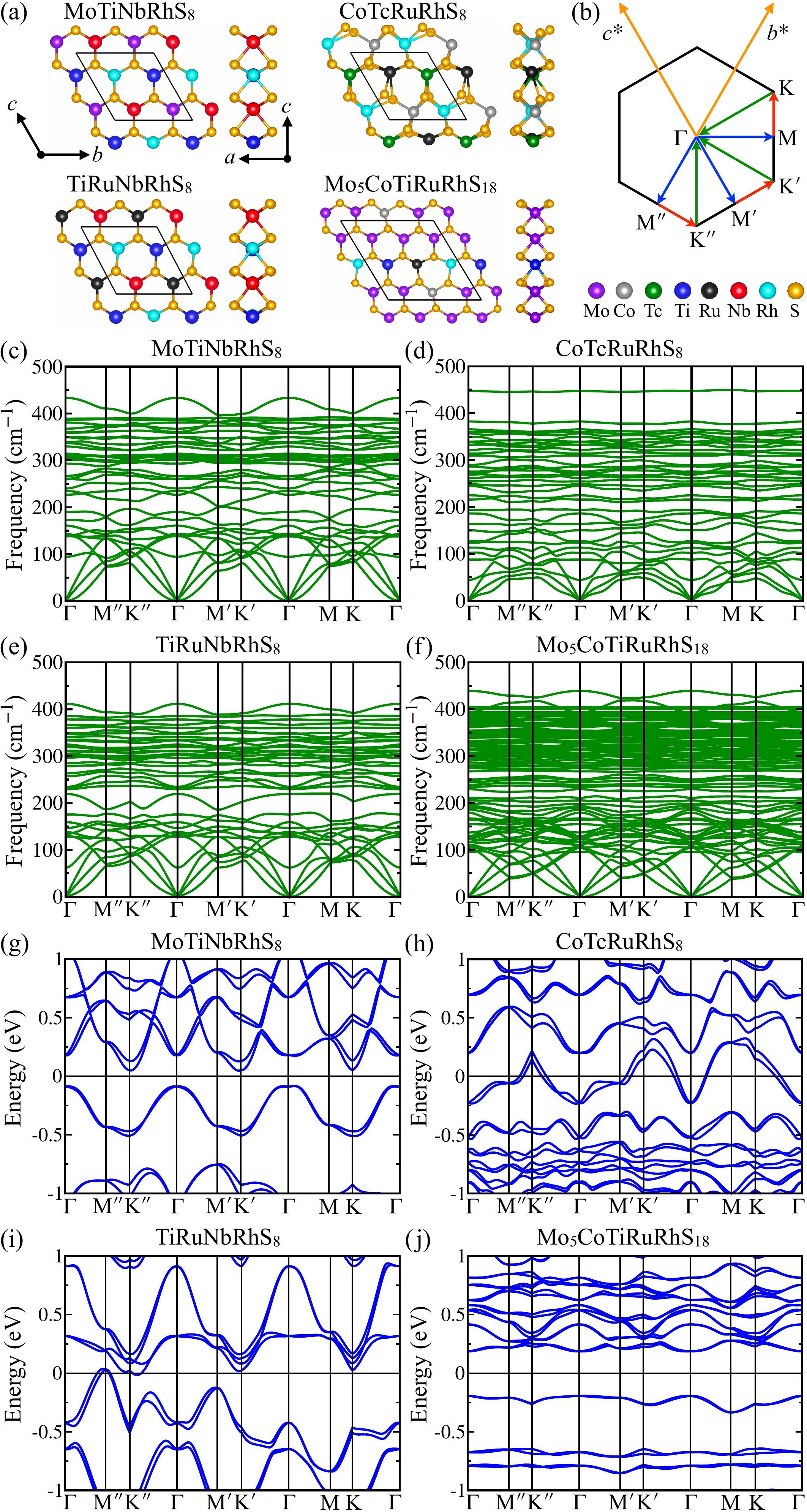}
\caption{(a) Structures of MoTiNbRhS$_8$, CoTcRuRhS$_8$, TiRuNbRhS$_8$, and Mo$_5$CoTiRuRhS$_{18}$ and (b) the Brillouin zone. 
The phonon dispersions of (c) MoTiNbRhS$_8$, (d) CoTcRuRhS$_8$, (e) TiRuNbRhS$_8$, and (f) Mo$_5$CoTiRuRhS$_{18}$ along the paths 
shown in (b). The electronic band structures with the inclusion of spin-orbit coupling along the same paths are shown in (g), (h), (i), and (j), respectively.
}
\label{fig:Fig3}
\end{figure}

Here, we extend the (2$\times$2) supercell to the (3$\times$3) supercell and investigate two examples with more significant mixing: Mo$_5$CoTiRuRhS$_{18}$ and 
Mo$_3$CoTcTiRuNbRhS$_{18}$.  We aim to determine if the mixing-stabilized phonon instability can still be identified in a larger supercell. According to 
Table~\ref{table:totalenergy}, the compound Mo$_3$CoTcTiRuNbRhS$_{18}$, containing all the elements Mo, Co, Tc, Ti, Ru, Nb, and Rh, exhibits a lower total energy 
compared to both the total-energy sums obtained from the (1×1) and the (2×2) XS$_2$ compounds. On the other hand, Mo$_5$CoTiRuRhS$_{18}$ only has a lower total energy 
than the sum of the corresponding (1$\times$1) XS$_2$ compounds. However, both compounds are dynamically stable without exhibiting imaginary-frequency branches 
in the phonon dispersion. The structure and phonon dispersion of Mo$_5$CoTiRuRhS$_{18}$ are shown in Figs.~\ref{fig:Fig3} (a) and (f), respectively, and 
the results for Mo$_3$CoTcTiRuNbRhS$_{18}$ can be found in Supplemental Material\cite{Supplement}. The findings suggest that more stable high-entropy XS$_2$ compounds, 
where X represents a highly mixed combination of Mo, Co, Tc, Ti, Ru, Nb, and Rh, may be realized in larger supercells and gain energy at the level of 
total energy.

Finally, concerning the applications for next-generation devices, the high-entropy semiconductors are expected 
to be realized in the investigated XS$_2$ compounds through the mixing of transition metal elements with different supercell sizes 
or even without a short range of periodicity. The high-entropy XS$_2$ compounds offer a variety of band 
structures with different gap sizes and types, such as indirect-gap semiconductors, metals, semimetals, and direct-gap semiconductors, as illustrated in 
Figs.~\ref{fig:Fig3} (g)-(j) by MoTiNbRhS$_8$, CoTcRuRhS$_8$, TiRuNbRhS$_8$, and Mo$_5$CoTiRuRhS$_{18}$, respectively. The band-gap types, which describe the 
electrical conductivity properties, of all the studied compounds are listed in Table~\ref{table:totalenergy}. It's worth mentioning that even though 
(2$\times$2) RuS$_2$ and (2$\times$2) RhS$_2$ are categorized as indirect-gap semiconductors, the band structures suggest 
that they are close to direct-gap semiconductors, indicating the possible tunability through, for example, external strain. Furthermore, the band structure of 
MoRuNbRhS$_8$ indicates that it is close to a zero-gap semimetal, and the sign of the gap can be easily tuned by strain and temperature.
The band structures of all the studied compounds can be found in Supplemental Material\cite{Supplement}.

In conclusion, mixing transition metal elements to create high-entropy combinations in the 1H-MoS$_2$ structure may eliminate the phonon instability dictated 
in CoS$_2$, TcS$_2$, TiS$_2$, RuS$_2$, NbS$_2$, and RhS$_2$ and gain energy at the level of total energy without the help of the TS term in the Gibbs free energy. 
The high-entropy XS$_2$ compounds do not really need the metal-insulator transitions and large atomic distortions to stabilize the systems in general. 
This is because the different atomic species already provide, to some extent, the effect of structural distortions in the charge density waves.
The illustrated cases obtained in the (2$\times$2) and (3$\times$3) supercells exhibit a rich set of band structures, including the direct- and indirect-gap 
semiconductors, metals, and semimetals, useful for electronic, optoelectronic, and other industrial applications.

\begin{acknowledgments}
The calculations were carried out using the facilities in JAIST and Tamkang University.
C.-C.L. thanks Professor Giulia Galli for a fruitful discussion. 
C.-C.L. acknowledges the National Science and Technology Council of Taiwan for financial support under Contract No. NSTC 112-2112-M-032-010 
and the support from the National Center for Theoretical Sciences (NCTS) of Taiwan.
Y. Y.-T. acknowledges support from JSPS KAKENHI Grant Numbers 21H05232 and 21H05236.
\end{acknowledgments}

\bibliography{refs}

\end{document}